\documentclass[journal,twoside]{IEEEtran}
\usepackage{mathtools}
\usepackage{amssymb}
\usepackage{cite}
\usepackage{bbm}
\usepackage{comment}

\newtheorem{lemma}{Lemma}
\newtheorem{definition}{Definition}
\newtheorem{remark}{Remark}
\newtheorem{example}{Example}

\DeclareMathOperator{\Si}{Si}
\DeclareMathOperator{\sinc}{sinc}
\DeclareMathOperator{\re}{Re}
\DeclareMathOperator{\im}{Im}
\DeclareMathOperator*{\argmin}{arg\,min}
\DeclareMathOperator*{\argmax}{arg\,max}

\begin{document}
\title{Why Constant-Composition Codes\\Reduce Nonlinear Interference Noise} 

\author{Reza~Rafie~Borujeny,~\IEEEmembership{Graduate~Student~Member,~IEEE}\\and~Frank~R.~Kschischang,~\IEEEmembership{Fellow,~IEEE}%
\thanks{Submitted for publication on March 4, 2022.  The authors are with the
Edward S. Rogers Sr. Department of Electrical and Computer Engineering,
University of Toronto, Toronto, ON M5S 3G4 Canada
(e-mail:\{rrafie,frank\}@ece.utoronto.ca).}}

\maketitle

\begin{abstract}
A time-domain perturbation model of the nonlinear Schr\"{o}dinger equation is
used to study wave-length-division multiplexed communication over a single
polarization. The model explains (a) why constant-composition codes offer an
improvement in signal to noise ratio compared with independent and uniform
selection of constellation points and (b) why similar gains are obtained using
carrier recovery algorithms even without using constant-composition codes.
\end{abstract}


\section{Introduction}

\IEEEPARstart{I}{nterchannel} interference induced by the effects of Kerr nonlinearity is a major
factor limiting the information-carrying capacity of
wave-length-division-multiplexed (WDM) optical transport networks.  At
transmission power levels beyond a certain optimum, the harmful effects of
nonlinear signal--signal interactions resulting in cross-phase modulation (XPM)
outweigh the benefits of transmitting a stronger signal
\cite{essiambre2010capacity, mecozzi2012nonlinear}.  Designing communication
schemes that can tolerate or compensate XPM is under active
research~\cite{dar2017nonlinear}.

Recent studies on probabilistic shaping for fiber optic systems have reported
that a considerable gain in signal-to-noise ratio (SNR) can be achieved in the
presence of Kerr nonlinearity using distribution matchers of short blocklength,
but that the gain reduces (or is absent) at longer blocklength.
In~\cite{amari2019introducing}, it is shown that finite blocklength enumerative
sphere shaping and constant-composition (CC) distribution matching achieve
higher effective SNRs than their infinite blocklength counterparts.  The same
observation is made in~\cite{fehenberger2020mitigating}.
In~\cite{fehenberger2020analysis}, the effect of a CC distribution matcher on
the induced nonlinear interference is attributed to a limited concentration of
identical symbols. The effect of blocklength of a distribution matcher on the
nonlinear interference noise is further studied in~\cite{peng2021baud}.
In~\cite{wu2021temporal}, the temporal energy behavior of symbol sequences is
studied and a new metric, called energy dispersion index, is proposed to predict
the impact of blocklength of the distribution matcher on the effective SNR for
systems that use a CC distribution matcher for probabilistic shaping.  Similar
observations, but in a different context, have been reported as early as the
work in~\cite{wickham2004bit}, where it is shown that to estimate the nonlinear
interference accurately, longer bit sequences should be simulated, since
simulations with shorter bit sequences induce higher SNRs.

While probabilistic shaping with finite-length CC codebooks provides
demonstrable improvements in effective SNR in the absence of phase-tracking
algorithms at the receiver, the authors of~\cite{civelli2020interplay} show
that, in the nonlinear regime of operation, the gain in generalized mutual
information obtained by probabilistic amplitude shaping via enumerative sphere
shaping is the same as the gain of typical carrier phase recovery algorithms.
In particular, no additional shaping gain is observed when a carrier phase
recovery module is in place, which is the case in all practical systems. This
observation raises the question of whether or not the SNR gain of short-length
distribution matchers is of practical importance.

%Recently, the authors of~\cite{secondini2021new} consider shaping by sequence
%selection. Based on the induced nonlinear interference noise obtained by
%simulated transmission of candidate sequences of a given length, those with the
%lowest nonlinear interference energy are selected to form a code. Achievable
%information rates are calculated using various detection methods. Considerable
%increase in the achievable information rates are obtained compared with the
%case of the nonlinear channel with ideal dispersion compensation and
%independent and identically distributed Gaussian input symbols with a detection
%metric suitable for an additive white Gaussian channel. As it turns out,
%sequence selection based on simulated single-channel noiseless transmissions is
%enough to obtain such gains.

In this paper, we provide a theoretical justification for the superior
performance (in terms of SNR) of short CC codes.  Our work is based on the the
first order perturbation model of~\cite{mecozzi2012nonlinear}.  We determine
computationally feasible expressions for the perturbation coefficients of the
most dominant XPM terms, and we verify that the dominant XPM coefficients are
slowly varying functions of the time-separation of symbols in neighboring
channels.   Our results show that a more general class of codes, namely
\emph{spherical codes}, have the potential to offer the same gain in SNR as
long as their blocklength is not too long.  We also explain why carrier phase
recovery algorithms can provide similar gains.

%Similar calculations are done for the most dominant self-phase modulation (SPM)
%coefficients. This will allow us to formulate the sequence selection procedure
%of~\cite{secondini2021new} using a much simpler channel model compared with the
%split-step model of~\cite{secondini2021new}.

The rest of the paper is organized as follows. In
Section~\ref{sec:channel_model}, the theory of the first order perturbation
method of~\cite{mecozzi2012nonlinear} is reviewed. This section includes a new
and efficient method to accurately calculate the most dominant perturbation
coefficients. Section~\ref{sec:cc_codes} reviews the main properties of CC
codes.  In Section~\ref{sec:justification}, we use the perturbation model to
explain why CC codes offer an SNR gain, and why carrier phase recovery can
provide similar gains.  Section~\ref{sec:conclusions} concludes the paper.

\section{A Mathematical Model for Nonlinear Signal--Signal Interactions}\label{sec:channel_model}

In this section, we re-derive the time domain perturbation model
of~\cite{mecozzi2012nonlinear}. This model will prove useful in the following
sections to investigate the nonlinear interference noise of CC codes.

\subsection{Channel Model}
Propagation of a narrow-band optical signal over a single mode fiber without
attenuation is described by the nonlinear Schr\"{o}dinger (NLS)
equation~\cite{govind2017nonlinear}
\begin{equation*}
    \frac{\partial Q(\tau,l)}{\partial l} = -i\frac{\beta_2}{2}\frac{\partial^2 Q(\tau,l)}{\partial \tau^2} + i\gamma \left| Q(\tau,l) \right| ^2 Q(\tau,l).
\end{equation*}
In this equation, the constant $\beta_2$ is the chromatic dispersion coefficient
and $\gamma$ is the nonlinearity coefficient. We assume $\beta_2<0$ which
corresponds to the anomalous-dispersion regime.  The assumption of having no
attenuation is merely to be able to isolate the effect of XPM in the following
sections. A more realistic loss profile or lumped amplification, along with
higher order dispersive effects can also be considered.

We work with a normalized NLS equation by considering the following change of
variables:
\begin{equation}\label{change_of_var}
q(z,t) = \frac{Q(l, \tau)}{\sqrt{P_0}}, \qquad z = -\frac{l}{L_0}, \qquad  t = \frac{\tau}{T_0}
\end{equation}
where $L_0$ is a free parameter, $T_0 = \sqrt{\left| \beta_2\right|L_0/2}$, and
$P_0 = \frac{2}{\gamma L_0}$.  Using (\ref{change_of_var}), the NLS equation
becomes
\begin{equation}\label{eq:nls}
\mathcal{D}_zq = -i\mathcal{D}_t^2q - 2i\left| q\right|^2 q
\end{equation}
in which $\mathcal{D}_z$ and $\mathcal{D}_{t}$ are the spatial and temporal
differentiation operators with respect to the normalized spatial variable $z$
and the normalized temporal variable $t$, respectively. 

We start by considering the dispersive equation
\begin{equation*}
\mathcal{D}_z q = -i\mathcal{D}_t^2q
\end{equation*}
and consider it as our \emph{unperturbed} equation. We perturb this equation by
coupling the cubic nonlinearity as
\begin{equation}\label{eq:perturbed}
\mathcal{D}_z q = -i\mathcal{D}_t^2q -2i\epsilon\lvert q\rvert^2 q.
\end{equation}
The solution to this equation, which is a function of the parameter $\epsilon$,
may be written as $q(z,t;\epsilon)$.  We assume this solution admits a power
series representation in the coupling parameter $\epsilon$
\begin{equation}\label{eq:power_series}
q(z,t;\epsilon) = \sum_{m=0}^{\infty} \epsilon^m q_m(z,t).
\end{equation}
Later on, we set $\epsilon = 1$ to recover (\ref{eq:nls}).

We substitute (\ref{eq:power_series}) into (\ref{eq:perturbed}) to obtain the
dynamics of each $q_m$:
\begin{align*}
\sum_{m = 0}^{\infty}\epsilon^m \mathcal{D}_z q_m  = &-i\sum_{m = 0}^{\infty} \epsilon^m \mathcal{D}_t^2 q_m \\&-2i\sum_{m = 0}^{\infty}
\sum_{l = 0}^{m}\sum_{k = 0}^{m-l} \epsilon^{m+1} q_l q_k^*q_{m-l-k}.
\end{align*}
By equating the coefficients of the corresponding powers of $\epsilon$ on both
sides, we get
\begin{equation}
\mathcal{D}_zq_0 = -i\mathcal{D}_t^2q_0 
\end{equation}
and
\begin{equation}\label{eq:generic_m}
\mathcal{D}_zq_m = -i\mathcal{D}_t^2q_m - 2i\sum_{l=0}^{m-1}\left(\sum_{k=0}^{m-l-1} q_l q_k^*q_{m-l-k-1}\right)
\end{equation}
for $m = 1, 2, \hdots$. 
In the special case of $m = 1$, we get
\begin{equation}
\mathcal{D}_z q_1 = -i\mathcal{D}_t^2q_1 - 2i\lvert q_0\rvert^2q_0.
\end{equation}
We consider the channel model obtained by keeping terms in
(\ref{eq:power_series}) up to the first order in $\epsilon$.

To summarize, the channel model is given by one algebraic equation and two partial
differential equations, namely 
\begin{align}
q &= q_0 + \epsilon q_1,\nonumber \\
\mathcal{D}_z q_0 &= -i\mathcal{D}_t^2 q_0,\label{eq:eq2}\\
\mathcal{D}_z q_1 &= -i\mathcal{D}_t^2 q_1- 2i\lvert q_0\rvert^2 q_0,\label{eq:eq3}
\end{align}
together with the boundary conditions
\begin{align}
q_0(0, t) &= q(0, t),\nonumber \\
q_1(0, t) &= 0.\label{eq:bc2}
\end{align}
Recall that we eventually let $\epsilon = 1$ so that we recover (\ref{eq:nls}).

To study this channel model---and in general to study equations of the form
given in (\ref{eq:generic_m})---the following lemma is useful.
\begin{lemma}\label{lemma:generic_pde}
Let $f(z, t)$ be a sufficiently smooth function
that satisfies the partial differential
equation
\begin{equation}
\mathcal{D}_z f(z, t) = k\mathcal{D}_t^2 f(z, t) + g(z,t),
\end{equation}
together with the boundary condition
\begin{equation}
f(0, t) = \phi(t).
\end{equation}
Then
\begin{equation}
f(z,t) = \mathcal{D}\left[\phi(t);z, k\right] + \int_0^z \mathcal{D}\left[g(z', t);z-z', k\right]\,\mathit{dz'}
\end{equation}
with
\begin{equation}
\mathcal{D}[\cdot; z, k] \vcentcolon= \mathcal{F}^{-1}\left[e^{-k(2\pi f)^2 z}\mathcal{F}\left[\cdot\right]\right]
\end{equation}
in which $\mathcal{F}[\cdot]$ is the Fourier transform operator.
\end{lemma}

In our case of interest, we fix $k = -i$ in Lemma~\ref{lemma:generic_pde} and
drop the $k$-dependence of the dispersion operator~$\mathcal{D}$.  We summarize
some properties of the dispersion operator in
Lemmas~\ref{lemma:dispersion_impulse_response}--\ref{lemma:unitarity}.
\begin{lemma}\label{lemma:dispersion_impulse_response}
The impulse response of the dispersion operator is
\begin{equation*}
\mathcal{D}\left[\delta(t); z\right] = \frac{1}{\sqrt{4\pi z}}e^{-i\left(\frac{t^2}{4 z} - \frac{\pi}{4}\right)}.
\end{equation*}
When $z<0$, the square root should be interpreted as 
\begin{equation}
\sqrt{z} = i\sqrt{\lvert z\rvert}. 
\end{equation}
\end{lemma}
\begin{lemma}\label{lemma:dispersion_modulation}
If 
\[
\mathcal{D}\left[s(t); z\right] = S(z, t)
\]
then
\[
\mathcal{D}\left[s(t-t_0)e^{i2\pi f_0t}; z\right] = S(z, t  - t_0 + 4\pi f_0 z)e^{i2\pi f_0\left(t + 2\pi f_0 z\right)}.
\]    
\end{lemma}
\begin{lemma}\label{lemma:unitarity} The dispersion operator is unitary, that is,
\[
\int_{-\infty}^{\infty}\mathcal{D}\left[s_1(t); z\right] \mathcal{D}^*\left[s_2(t); z\right]\,\mathit{dt} = \int_{-\infty}^{\infty} s_1(t) s_2^*(t)\,\mathit{dt}.
\]
\end{lemma}

Applying Lemma~\ref{lemma:generic_pde} to (\ref{eq:eq2}), we get
\begin{equation}\label{eq:eq12}
q_0(z, t) = \mathcal{D}\left[q(0, t); z\right].
\end{equation} 
Considering the boundary condition (\ref{eq:bc2}), if we apply Lemma~\ref{lemma:generic_pde} to (\ref{eq:eq3}) we get
\begin{equation}\label{eq:eq22}
q_1(z, t) = -2i\int_0^z \mathcal{D}\left[\lvert q_0(z', t)\rvert^2q_0(z', t); z-z'\right]\,\mathit{dz'}.
\end{equation}
Notice that (\ref{eq:eq12}) and (\ref{eq:eq22}) give an alternative description
of the channel law. For convenience, we write this alternative description here
again:
\begin{align}
q(z, t) &= q_0(z, t) + \epsilon q_1(z, t),\nonumber\\
q_0(z, t) &= \mathcal{D}\left[q(0, t); z\right],\label{eq:integralform_simple}\\
q_1(z, t) &= -2i\!\int_0^z\! \mathcal{D}\left[\lvert \mathcal{D}\left[q(0, t); z'\right]\rvert^2 \mathcal{D}\left[q(0, t); z'\right]; z\!-\!z'\right]\mathit{dz'}.\nonumber
\end{align}

\subsection{Model of Nonlinear Interference}\label{sec:model_nli}

Consider a WDM system with $2M+1$ channels using unit-energy $\sinc(\cdot)$
for pulse shaping for all channels. Assume also that all channels
co-propagate.  The signal constellation used by
channel $k$ is $A_k \subset \mathbb{C}$. The $j$th time
symbol sent over channel $k$ is denoted as $a_{k,j} \in A_k$.
The launched signal, therefore, is
\[
q(0, t) = \frac{1}{\sqrt{T}}\sum_{k=-M}^M\sum_{j=-\infty}^{\infty} a_{k,j}\sinc\left(\frac{t - jT}{T}\right)e^{i2\pi kBt},
\]
where $T^{-1}$ is the baud rate and $B = T^{-1}$ is the channel spacing.
By properly choosing the free parameter $L_0$, we may assume $T = 1$
which simplifies the launched signal to
\[
q(0, t) = \sum_{k=-M}^M\sum_{j=-\infty}^{\infty} a_{k,j}\sinc\left({t - j}\right)e^{i2\pi kt}.
\]
A constant phase shift and time-center drift can be introduced for each
channel. Variations of pulse shape or channel spacing are also possible. We
assume no such shifts, drifts or variations for brevity.

The channel of interest for us is the middle channel indexed by $k = 0$ and the
symbol of interest is the one indexed by $j = 0$, that is, $a_{0,0}$.  Signal
detection for the channel of interest is done by using a matched filter, i.e.,
a $\sinc$ function dispersively propagated to distance $z$.  Thus, for a
fixed $z$, the matched-filter impulse response $h(t)$ is described by
\[
h^*(t) = \mathcal{D}\left[\sinc\left(t\right); z\right].
\]
Therefore, to detect $a_{0,0}$ at distance $z$, we look at
\begin{align}
\hat{a}_{0,0} &= \int_{-\infty}^{\infty} q(z, t) h(t)\,\mathit{dt} \nonumber \\
&= \int_{-\infty}^{\infty} q_0(z, t) h(t)\,\mathit{dt} + \epsilon\int_{-\infty}^{\infty} q_1(z, t) h(t)\,\mathit{dt}.\label{eq:mf2}
\end{align}
Using Lemma~\ref{lemma:unitarity}, the first term in (\ref{eq:mf2}) becomes
$a_{0,0}$.  The second term in (\ref{eq:mf2}) is responsible for nonlinear
signal--signal interactions. As is shown in~\cite{essiambre2010capacity}, near
the ``peak'' of the achievable information rate curve, the signal--signal
interchannel nonlinearities are responsible for limiting the capacity in the
WDM systems considered. To simplify the expressions that follow,
we introduce the shortened notation
\begin{equation}\label{eq:def_D}
D(k, j, z, t) \vcentcolon= \mathcal{D}\left[\sinc\left({t - j}\right)e^{i2\pi kt}; z\right].
\end{equation}
Lemma~\ref{lemma:dispersion_modulation} may be used when computing (\ref{eq:def_D}).
We also define 
\begin{align}\label{eq:littlec}
&c(k_1, k_2, k_3, j_1, j_2, j_3, z, t) \vcentcolon= \\
&D(k_1, j_1, z, t)D^*(k_2, j_2, z, t)D(k_3, j_3, z, t)D^*(0, 0, z, t) \nonumber
\end{align}
and
\begin{align}\label{eq:C}
&C(k_1, j_1, k_2, j_2, k_3, j_3, z)\vcentcolon= \nonumber\\
&-2i\int_{0}^{z}\int_{-\infty}^{\infty} c(k_1, k_2, k_3, j_1, j_2, j_3, z', t)\,\mathit{dt}\mathit{dz'}
\end{align}
to study the second term in (\ref{eq:mf2}) when the channel is described by (\ref{eq:integralform_simple}).
Denoting the integral in the second term of (\ref{eq:mf2}) as $\Delta$, by
using Lemma~\ref{lemma:unitarity},
\begin{align}\label{eq:second_term}
\Delta \vcentcolon=& \int_{-\infty}^{\infty} q_1(z, t) h(t)\,\mathit{dt} \\
 =&\sum_{k_1,k_2,k_3}\sum_{j_1,j_2,j_3}\!\!\! a_{k_1,j_1}a_{k_2,j_2}^* a_{k_3, j_3} C(k_1, j_1, k_2, j_2, k_3, j_3, z).\nonumber
\end{align}
Multiplication by $D^*(0, 0, z', t)$ in (\ref{eq:littlec}) and (\ref{eq:C})
can be seen as low-pass filtering. Accordingly, one can show that if
\begin{equation*}
\lvert k_1 - k_2 + k_3\rvert > 1
\end{equation*}
then
\begin{equation*}
C(k_1, j_1, k_2, j_2, k_3, j_3, z) = 0.
\end{equation*}
As a result, for $2M+1$ WDM channels, there are exactly
\begin{equation*}
9M^2 + 9M + 1
\end{equation*}
nonzero $C(\cdot, j_1, \cdot, j_2, \cdot, j_3, z)$ terms.
For example, with 5 channels, we have 
$M = 2$
and the number of such nonzero terms is 55. One should note the extra symmetry
\begin{equation*}
C(k_1, j_1, k_2, j_2, k_3, j_3, z) = C(k_3, j_3, k_2, j_2, k_1, j_1, z)
\end{equation*}
to reduce the complexity of computing these nonzero terms.

The dominant terms in (\ref{eq:second_term}) are those with 
\[
k_1 = k_2 = k_3 = 0,
\]
which correspond to self-phase modulation (SPM), and those with
\[
k_1 = k_2, k_3=0,
\]
\[
k_1 = 0, k_2 = k_3,
\]
which represent XPM. Hence,
\[
\Delta \approx \Delta_{\text{SPM}} + \Delta_{\text{XPM}}
\]
where
\begin{equation*}
\Delta_{\text{SPM}} = \sum_{j_1,j_2,j_3} a_{0,j_1}a_{0,j_2}^* a_{0,j_3} C(0, j_1, 0, j_2, 0, j_3, z)
\end{equation*}
and
\begin{equation}\label{eq:xpm_all}
\Delta_{\text{XPM}} = 2\sum_{k_1\neq0}\sum_{j_1,j_2,j_3}\!\!\! a_{k_1,j_1}a_{k_1,j_2}^* a_{0,j_3} C(k_1, j_1, k_1, j_1, 0, j_3, z).
\end{equation}
Of the terms in (\ref{eq:xpm_all}), the most dominant ones are those with
\[
j_1 = j_2, j_3 = 0.
\]
We define the new notation $\chi_{k,j}(z)$ to describe such terms:
\[
\chi_{k,j}(z) \coloneqq C(k, j, k, j, 0, 0, z).
\]
Consequently, we approximate XPM by
%\begin{equation}\label{eq:spm}
%\Delta_{\text{SPM}} \approx 2a_{0,0}\sum_{j\neq0} \lvert a_{0,j}\rvert^2 \chi_{0,j}(z) + a_{0,0}\lvert a_{0,0}\rvert^2\chi_{0,0}(z).
%\end{equation}
\begin{equation}\label{eq:xpm}
\Delta_{\text{XPM}} \approx 2a_{0,0}\sum_{k\neq 0}\sum_{j} \lvert a_{k,j}\rvert^2 \chi_{k,j}(z).
\end{equation}
Naturally, calculation of the coefficients $\chi_{k,j}(z)$ becomes important.
From Lemma~\ref{lemma:dispersion_impulse_response}, the impulse response of the
dispersion operator is an even function of time. One implication of this is
that $D(0,0,z,t)$ is always an even function of time.
Using this observation and Lemma~\ref{lemma:dispersion_modulation}, one can
show that
\[
\chi_{k,j}(z) = \chi_{-k,-j}(z).
\]
This extra symmetry will be helpful when calculating the perturbation coefficients.

%\footnote{This observation is intuitive! Just change the observer from $a_{0,0}$ to $a_{-k,-j}$.}
%%%%%%%%%%%%%%%%%%%%%%%%%%%%%%%%%%%%%%%%%%%
\begin{figure}[t]
\centering
\includegraphics[width=\columnwidth]{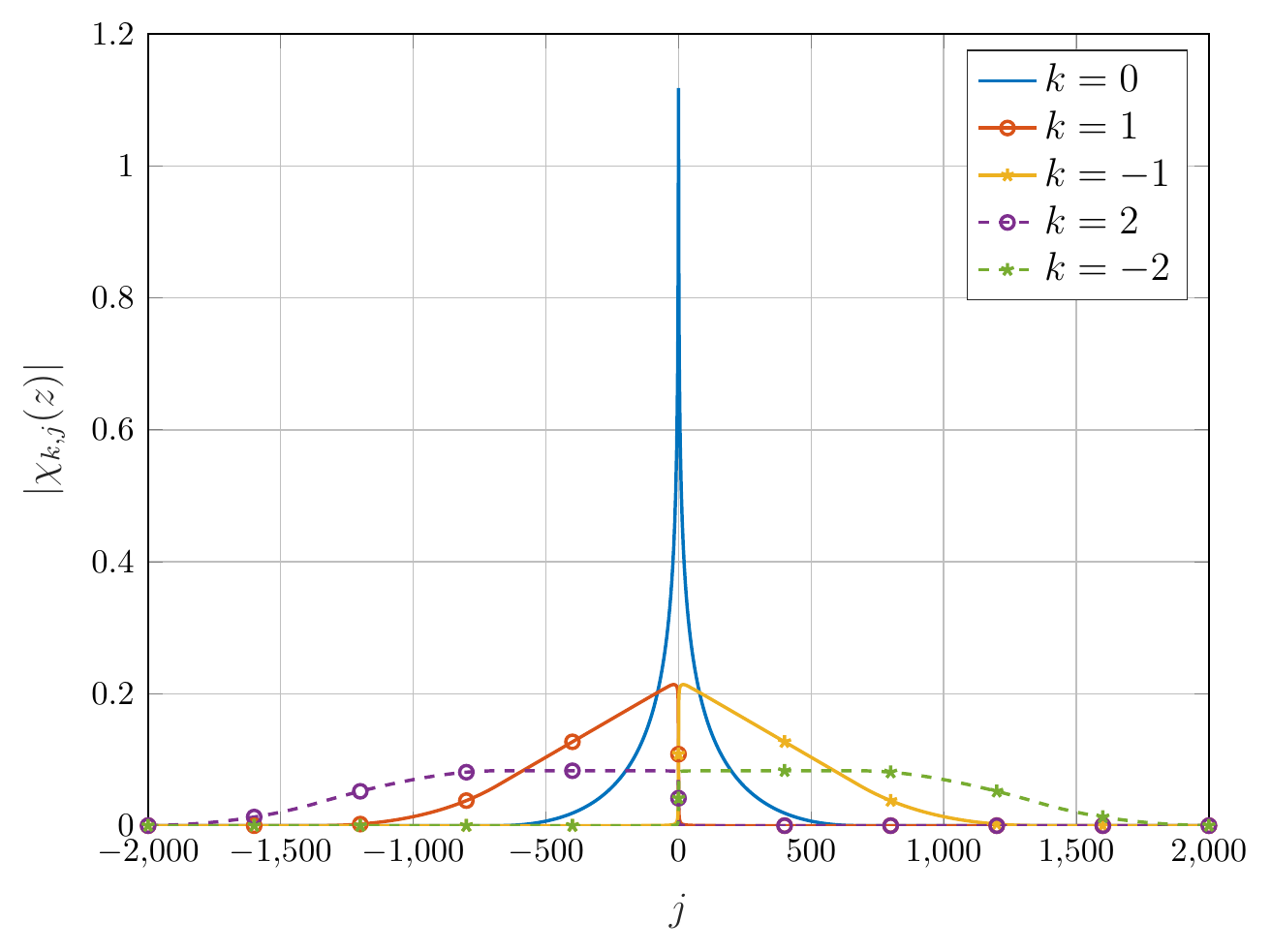}
\caption{The perturbation coefficients for a fiber of length $2000$~km with $B = 50$~GHz.}
\label{fig:c_2000}
\end{figure}
%%%%%%%%%%%%%%%%%%%%%%%%%%%%%%%%%%%%%%%%%%%

Using Rayleigh's energy theorem and (\ref{eq:C}), we get the following lemma.
\begin{lemma}\label{lemma:C_k_j}
The coefficient $\chi_{k,j}(z)$ can be found by evaluating the following integral
\begin{equation}\nonumber
-2i\int_{k-1}^{k+1}\left(1-\lvert f-k \rvert\right)^2s(j,k,z,f)\,\mathit{df}
\end{equation}
where
\begin{equation}\label{eq:spatial_integral}
s(j,k,z,f) = \int_{0}^{z} \sinc^2\left((j-4\pi fz')(1-\lvert f-k \rvert)\right)\,\mathit{dz'}.
\end{equation}
\end{lemma}

%%%%%%%%%%%%%%%%%%%%%%%%%%%%%%%%%%%%%%%%%%%
\begin{figure}[t]
\centering
\includegraphics[width=\columnwidth]{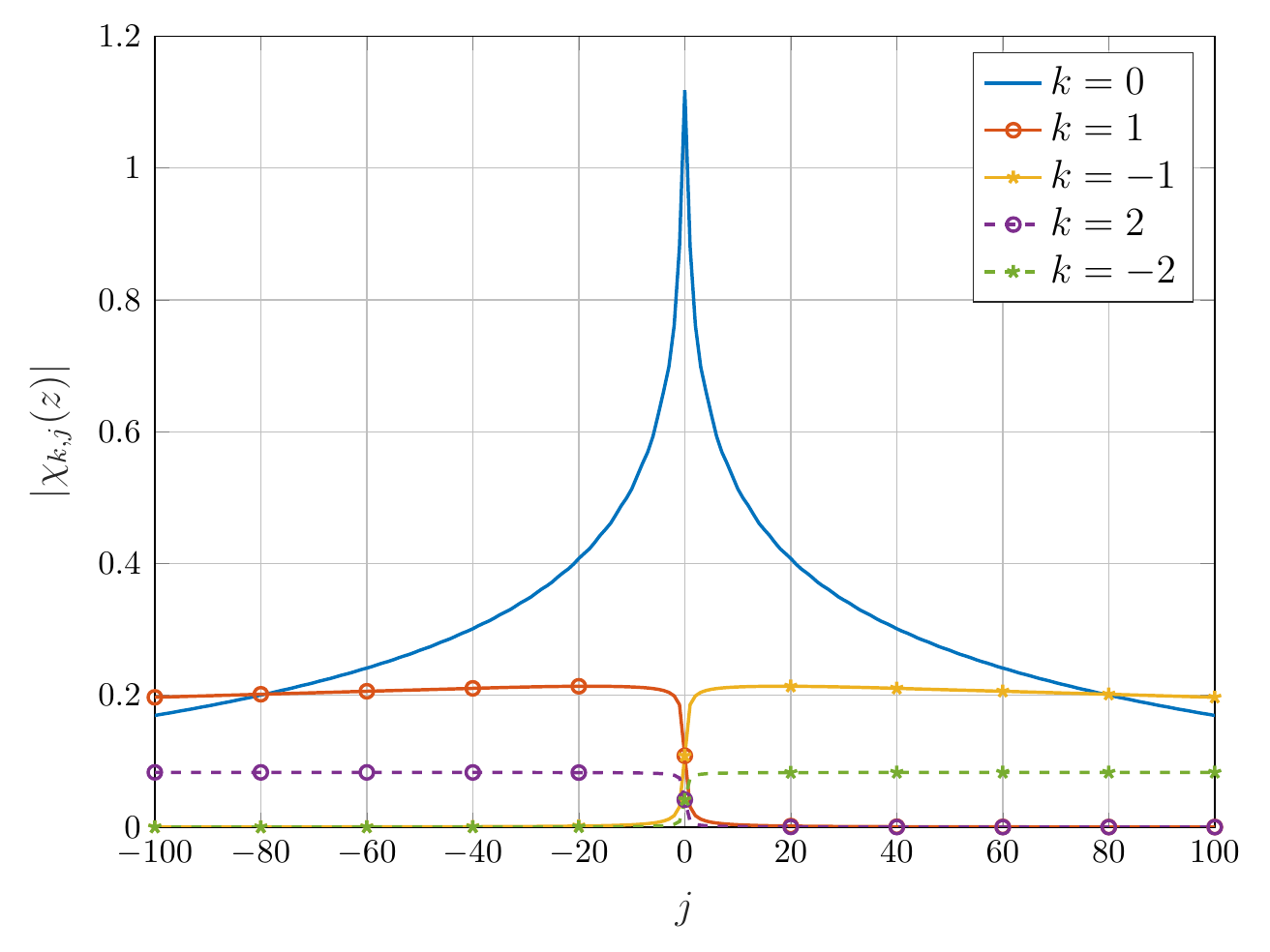}
\caption{The perturbation coefficients for a fiber of length $2000$~km with $B = 50$~GHz.}
\label{fig:c_100}
\end{figure} 
%%%%%%%%%%%%%%%%%%%%%%%%%%%%%%%%%%%%%%%%%%%

The spatial integral (\ref{eq:spatial_integral}) in Lemma~\ref{lemma:C_k_j} is
related to the sine integral $\Si(x)$ defined as
\[
\Si(x) = \int_0^x\frac{\sin t}{t}\,\mathit{dt},
\]
which can be efficiently calculated using a Pad\'{e}
approximation~\cite{rowe2015galsim} or other rational
approximations~\cite{macleod1996rational}.  Using the definition of the sine
integral, one can easily verify that
\begin{equation}
\int \sinc^2(x)\,\mathit{dx} = \frac{\Si(2\pi x)}{\pi} -x\sinc^2(x).
\end{equation}
As a result, we have
\begin{align*}
\int \sinc^2(ax+b)\,\mathit{dx} = &\frac{\Si(2\pi (ax+b))}{a\pi}\\
 &-\frac{(ax+b)}{a}\sinc^2(ax+b).
\end{align*}
By choosing
\[
a = -4\pi f(1-\lvert f - k\rvert),\quad b = j(1-\lvert f - k\rvert), \quad x = z',
\]
we find the antiderivative needed to evaluate the spatial integral of
Lemma~\ref{lemma:C_k_j}. The absolute value\footnote{Since $\chi_{k,j}(z)$
is purely imaginary, knowing the absolute value is enough to know the
coefficient.} of $\chi_{k,j}(z)$ for a fiber of length $2000$~km is shown in
Fig.~\ref{fig:c_2000} and Fig.~\ref{fig:c_100}. The channel spacing is $B =
50$~GHz. In particular, Fig.~\ref{fig:c_100} verifies that, as long as $j$ is
not close to zero, the XPM coefficients are indeed slowly varying functions of
$j$.

\section{Constant-Composition Codes}\label{sec:cc_codes}

For any positive integer $n$, the set of all possible
$n$-tuples drawn from any nonempty $m$-ary alphabet
$ A = \{a_1, a_2, \dots, a_m\}$ is denoted as $A^n$.
An $n$-\emph{permutation} is any invertible
function mapping $\{ 1, \ldots, n \}$ to itself.  The set of all
$n$-permutations forms a group under composition called the symmetric group
$S_n$.  The $n$-permutations
act on $n$-tuples over $A$ by a permutation of coordinates,
so that for any $p \in S_n$ and any $u = (u_1,\ldots,u_n) \in A^n$,
$p \cdot u = (u_{p(1)}, \ldots, u_{p(n)})$.

A code $C$ of length $n$ over $A$ is any nonempty subset of $A^n$.
The elements of $C$ are called codewords.
The rate, $R$, of a code $C$ is defined as
\[
R \coloneqq \frac{\log_2 \lvert C\rvert}{n}\quad \left[\frac{\text{bit}}{\text{symbol}}\right].
\]
A code $C$ is called a \emph{constant-composition code} if every
codeword in $C$ is a permutation of some fixed word $u \in A^n$.
More precisely, we have the following definition.
\begin{definition}\label{def:cc}
A constant-composition code $C$ of length $n$ over an alphabet $A$ is
a nonempty subset of $A^n$ such that for some $u\in A^n$ and
for some $P \subset S_n$,
\[
C = \{p \cdot u : p \in P \}.
\]
Such a code is denoted as a $(u,P)$ CC code.
\end{definition}

In this paper we consider only the special case where $P = S_n$.
Associated with each $n$-tuple
$v = (v_1, v_2, \dots, v_n) \in A^{n}$ over the $m$-ary alphabet
$A$ is an $m$-tuple called a \emph{type} which counts the number
of times each element of $A$ occurs as a coordinate of $v$.
In particular
\[
\mathrm{type}(v) = \left(w_1(v), w_2(v), \dots, w_m(v)\right),
\]
where
\[
w_k(v) = \sum_{j=1}^n\mathbbm{1}\left\{v_j = a_k\right\}, \quad k = 1, 2, \dots, m.
\]
The rate of the $(v,S_n)$ CC code as defined in Definition
\ref{def:cc} is then
\[
R = \frac{1}{n}\log_2 \frac{n!}{w_1(v)!w_2(v)!\cdots w_m(v)!},
\]
where the argument of the logarithm is the multinomial coefficient
that gives the number of distinct permutations of $v$.
In the rest of this paper, we focus on CC codes with 
\[
w_1 = w_2 = \dots = w_m = 1
\]
so that $m=n$ and
\[
R = \frac{1}{m}\log_2 m! \approx \log_2\frac{m}{e}.
\]

\begin{figure}[t]
\centering
\includegraphics[width=\columnwidth]{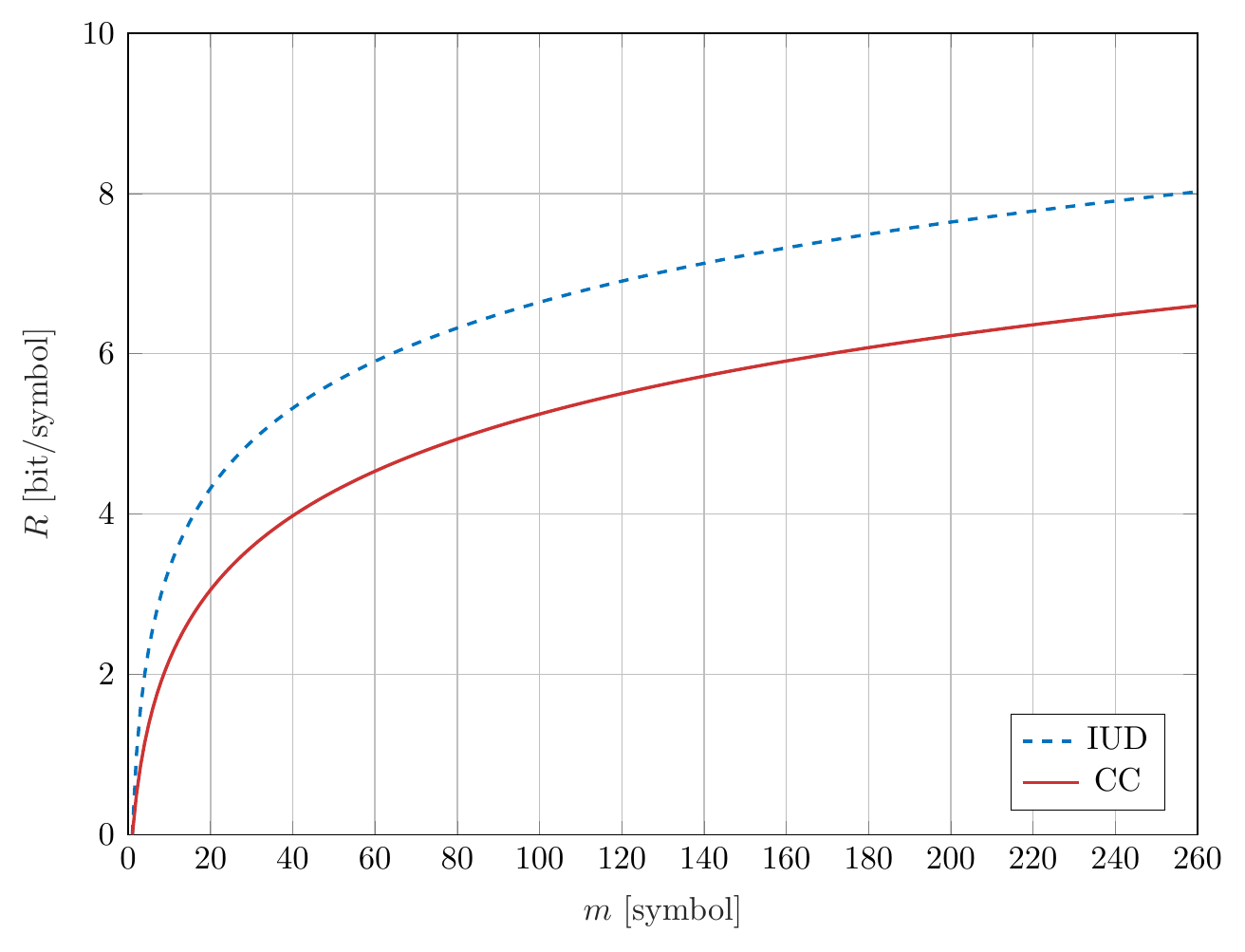}
\caption{Transmission rate of CC codes and IUD transmission.}
\label{fig:cc_rate}
\end{figure}

The alphabet $A$ that we consider is a subset of the complex numbers
$\mathbb{C}$. We will compare performance of such CC codes with a transmission
scheme where the symbols are selected independently and uniformly
at random from a quadrature amplitude modulation (QAM) constellation.
If the size of the constellation is $m$, the transmission rate
for an independent and uniformly distributed (IUD) selection of
points is
\[
R = \log_2 m \quad \left[\frac{\text{bit}}{\text{symbol}}\right].
\]
The rate of a CC code of length $m$ is compared with the transmission rate of
an IUD transmission scheme with a constellation of size $m$ in Fig.
\ref{fig:cc_rate}. Notice that the gap between the two curves converges to
\[
\log_2 e \approx 1.44.
\]

\begin{example}
When transmitting IUD from a QAM constellation, an alphabet of size $64$ gives
a transmission rate of $6$ bits per symbol, while to get the same transmission
rate using a CC code, an alphabet size of $171$ is needed.
\end{example}
%%%%%%%%%%%%%%%%%%%%%%%%%%%%%%%%%%%%%%%%%%%
%%%%%%%%%%%%%%%%%%%%%%%%%%%%%%%%%%%%%%%%%%%

\section{Cross-Phase Modulation Induced by Constant-Composition Codes}\label{sec:justification}

In this section, we consider a WDM system in which all channels use a CC code
of length $m$. We assume that the symbols of each codeword are transmitted
consecutively. If the $l^{\text{th}}$ codeword for transmission over channel
$k$ is 
\[
(a_{k, lm}, a_{k,lm+1}, a_{k,lm+2}, \dots, a_{k,lm+m-1}),
\]
the symbols that we send on channel $k$ from time $j=lm$ to time $j=lm+m-1$ are 
\[
a_{k, lm}, a_{k,lm+1}, a_{k,lm+2}, \dots, a_{k,lm+m-1}.
\]

As shown in Section~\ref{sec:model_nli}, the coefficients $\chi_{k,j}(z)$ are
slowly varying with $j$.  For example, for the parameters used in
Fig.~\ref{fig:c_2000} we have $\chi_{k,j}\approx\chi_{k,h}$ if
\[
\lvert j\rvert>10,\quad \lvert h\rvert>10 
\]
and
\[
\lvert j-h\rvert
\]
is not too large.  As long as the blocklength $m$ of the CC code used by all
channels is small enough so that the above approximation is justified, the XPM
term in (\ref{eq:xpm}) can be written as
\begin{align}
\Delta_{\text{XPM}} \approx & 2a_{0,0}\sum_{k\neq 0}\sum_{j=-m}^{m-1} \lvert a_{k,j}\rvert^2 \chi_{k,j}(z)\\
& +2a_{0,0}\sum_{k\neq 0}\sum_{l=1}^{\infty} \chi_{k,lm}(z)\sum_{j=0}^{m-1} \lvert a_{k,lm+j}\rvert^2\nonumber\\
& +2a_{0,0}\sum_{k\neq 0}\sum_{l=-\infty}^{-2} \chi_{k,lm}(z)\sum_{j=0}^{m-1} \lvert a_{k,lm+j}\rvert^2.\nonumber
\end{align}
If we denote the \emph{energy} of each of the codewords in the CC code used by
$E$, the XPM term becomes
\begin{align}\label{eq:xpm_approx}
\Delta_{\text{XPM}} \approx & 2a_{0,0}\sum_{k\neq 0}\sum_{j=-m}^{m-1} \lvert a_{k,j}\rvert^2 \chi_{k,j}(z)\\
& +2a_{0,0}E\sum_{k\neq 0}\sum_{l=1}^{\infty} \chi_{k,lm}(z)\nonumber\\
& +2a_{0,0}E\sum_{k\neq 0}\sum_{l=-\infty}^{-2} \chi_{k,lm}(z).\nonumber
\end{align}
Notice that the second and the third summations in (\ref{eq:xpm_approx}) are
deterministic. In other words, the most important XPM terms are those that
capture the nonlinear interaction of the symbol of interest with the symbols in
the neighboring channels that are closest in time to the symbol of interest at
the beginning of the fiber. This observation tells us that, as long as the
blocklength of the CC code is not too large, the effect of most of the XPM
terms in (\ref{eq:xpm}) is deterministic. The XPM uncertainty, therefore, will
be limited to the collision of the symbols that are transmitted almost
concurrently. Because of this limited XPM uncertainty, the overall observed SNR
will be higher than the case of IUD symbol selection.

\begin{remark}
In the above explanation, the only property of the CC codes that we have used
is the fact that CC codes are constant energy codes, i.e., all codewords of a
CC code have the same energy. As a result, we would expect the same kind of SNR
gains from more general constant energy codes. Constant energy codes are
usually known as spherical codes. While we do not intend to study the
performance of spherical codes in this paper, it would be interesting to see if
such codes can provide performance gains. This is especially important as
spherical codes can provide the same rates as CC codes with shorter
blocklengths.
\end{remark}

While CC codes are expected to reduce the uncertainty of XPM, we may at the
same time observe from our model that in general the XPM term induced in
detection of $a_{0,j}$ and $a_{0,h}$ is almost the same, provided that
$\lvert j-h\rvert$
is not too large.  In other words, we expect the effect of XPM on nearby
symbols to be nearly the same.  This observation, together with the fact that
XPM mostly affects the \emph{phase} of the detected symbols, suggest that most
of the XPM induced nonlinear interference may be undone by use of
phase-tracking algorithms as used in carrier recovery.   Furthermore, this
should be true even if IUD selection of symbols is being used.

%%%%%%%%%%%%%%%%%%%%%%%%%%%%%%%%%%%%%%%%%%%
\begin{figure}[t]
\centering
\includegraphics[width=\columnwidth]{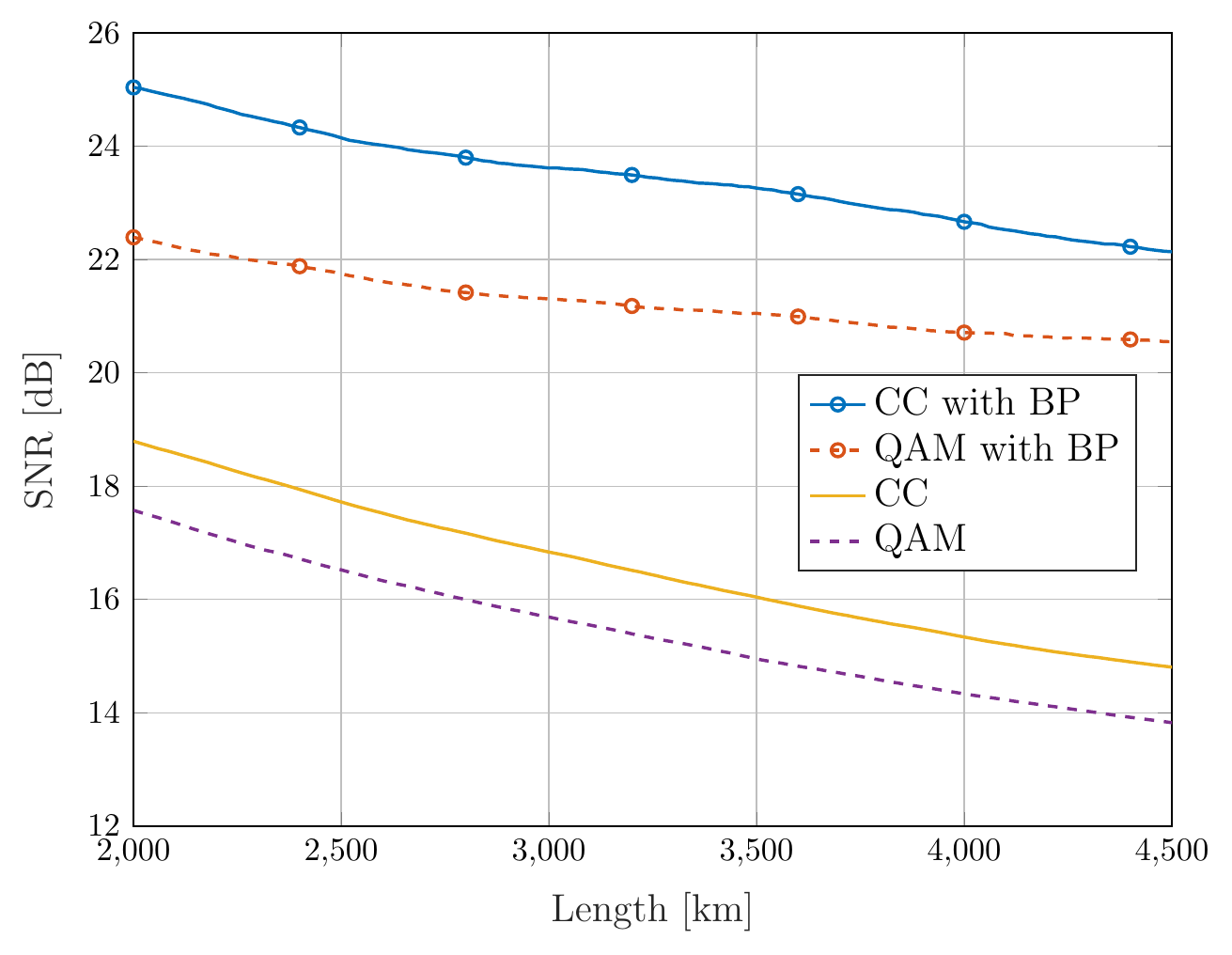}
\caption{SNR for a CC code of blocklength $171$ and IUD transmission using
$64$-QAM versus fiber length, in an idealized scenario
where there is no attenuation and no amplification noise, so that the nonlinear
interference noise is the only source of distortion. Two different detection
schemes are considered: back-propagation (BP) followed by
matched-filtering and matched-filtering without back-propagation.}
\label{fig:snr_nonoise}
\end{figure}
%%%%%%%%%%%%%%%%%%%%%%%%%%%%%%%%%%%%%%%%%%%

In the rest of this section, we verify these two observations by trying to
isolate the XPM noise in simulating data transmission using the split-step
Fourier method in two scenarios:  with and without carrier recovery.

\subsection{Experiments without Carrier Recovery}

To isolate the nonlinear interference, we consider simulating a WDM system
using the split-step Fourier method with adaptive step sizes
\cite{sinkin2003optimization} in an idealized setting.
The fiber is assumed to be lossless and there is
no amplification noise, thus allowing us to isolate and measure
interference noise.   Pulses are ideal $\sinc(\cdot)$ functions.
We assume that
five WDM channels travel along the fiber without any
adds and drops. The channel spacing is $50$~GHz.  No guard band is assumed.

Two types of detection are considered. The first one consists of detection
using a matched filter as explained in Section~\ref{sec:model_nli}.  In the
second detection method, the channel of interest is fully back-propagated after
being selection by a low-pass filter at the receiver. This is then followed by
matched filtering and sampling. Phase compensation is done by a
common phase rotation applied to all symbols chosen so 
that the average residual phase of the whole sequence of
symbols is $0$.

%%%%%%%%%%%%%%%%%%%%%%%%%%%%%%%%%%%%%%%%%%%
\begin{figure}[t]
\centering
\includegraphics[width=\columnwidth]{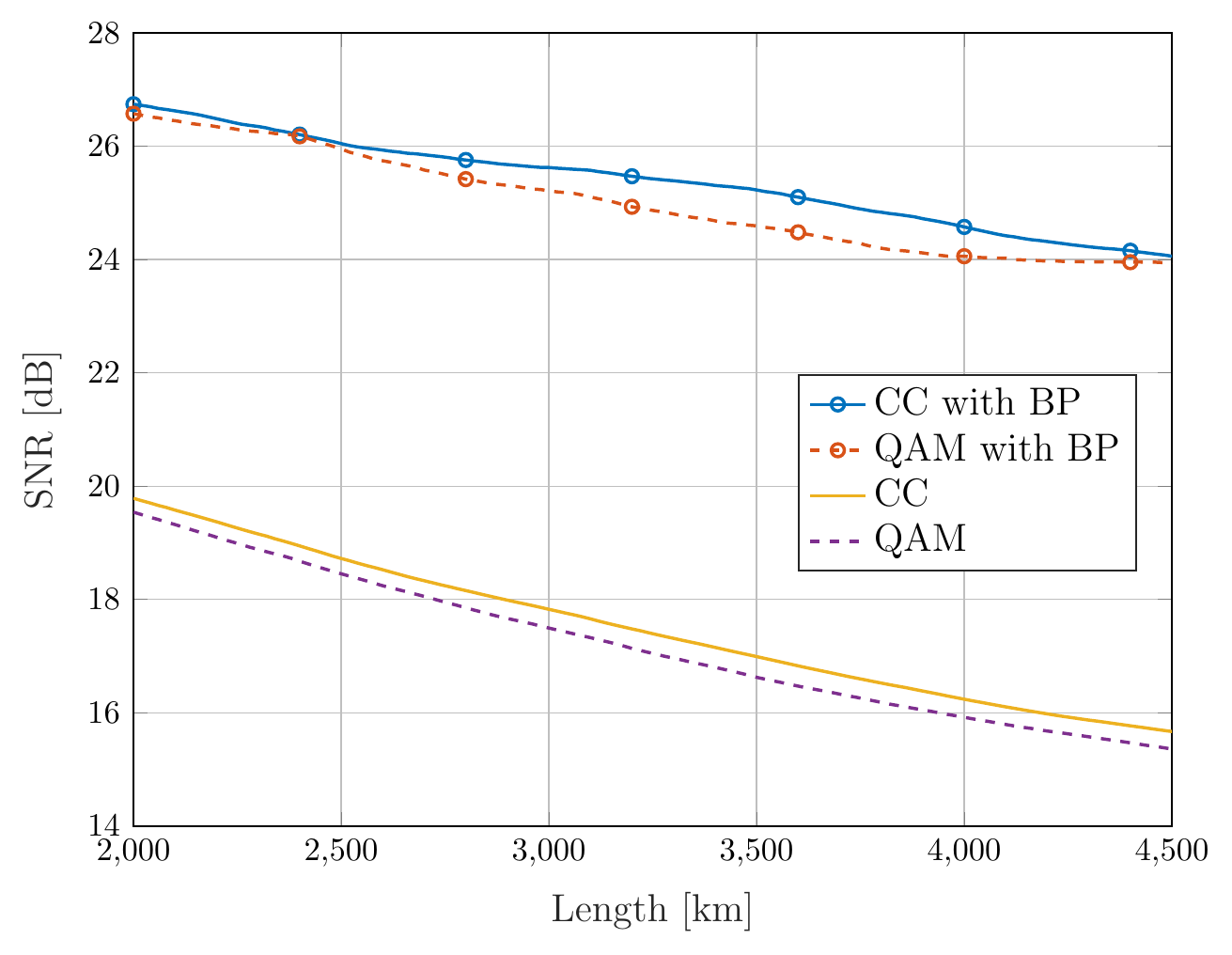}
\caption{SNR for a CC code of blocklength $171$ and IUD transmission using
$64$-QAM versus fiber length, in an idealized scenario where there is no
attenuation and no amplification noise, so that the nonlinear interference
noise is the only source of distortion.  Two different detection schemes are
considered: back-propagation (BP) followed by matched-filtering and
matched-filtering without back-propagation. The samples of the matched filter
are back-rotated by using a genie-aided blind phase search algorithm.}
\label{fig:snr_nonoise_bps}
\end{figure}
%%%%%%%%%%%%%%%%%%%%%%%%%%%%%%%%%%%%%%%%%%%

The resulting channel is modelled as an additive noise channel
\[
Y = X+N
\]
where $X$ is the input random variable, $Y$ is the output random variable, and $N$ is the additive noise (caused by interference) for which,
SNR is defined by
\[
\text{SNR } = \frac{\mathbb{E}[\lvert X\rvert^2]}{\mathbb{E}[\lvert N\rvert^2]}
\]
and is approximated by numerical averaging in place of statistical
expectations.  The transmission length has been varied from $2000$~km to
$4500$~km in steps of $20$~km. An IUD transmission with a square $64$-QAM, as
well as a CC code of blocklength $171$ are considered. The alphabet used for
the CC code is not optimized in any sense and is selected simply by picking
$171$ points of a $256$-QAM constellation having the least energy.

The results are shown in
Fig.~\ref{fig:snr_nonoise}. The results show that CC codes provide
a gain of about $1$~dB over IUD QAM without back-propagation, and they
provide a gain of about $2$~dB with back-propagation at the receiver.
Thus CC codes are indeed effective in reducing the uncertainty due
to XPM in the received signal.

\subsection{Experiments with Carrier Recovery}

Under the same setup, a different decoder that incorporates a blind phase
search (BPS) algorithm \cite{pfau2009hardware} at the very last step to
minimize the phase error is considered. The decoder used is a genie-aided one
as it uses the transmitted symbols to perform the best phase compensation that
one might expect from the blind phase search algorithm. When detecting the
symbol of interest $X_0 = a_{0,0}$, the BPS is done by solving the following
minimization problem:
\begin{equation}
\phi_0 = \argmin_{\theta} \sum_{j = -N}^{N}\lvert X_j - Y_j e^{i \theta}\rvert^2,
\end{equation}
in which $X_j$ is the $j^{\text{th}}$ transmitted symbol and $Y_j$ is the
corresponding output of the matched filter. The output of the matched filter is
rotated by $\phi_0$ and the output of the BPS block is 
$Y_0e^{i\phi_0}$.
The window size of the BPS block is set to $2N+1 = 21$.

%%%%%%%%%%%%%%%%%%%%%%%%%%%%%%%%%%%%%%%%%%%
\begin{figure}[t]
\centering
\includegraphics[width=\columnwidth]{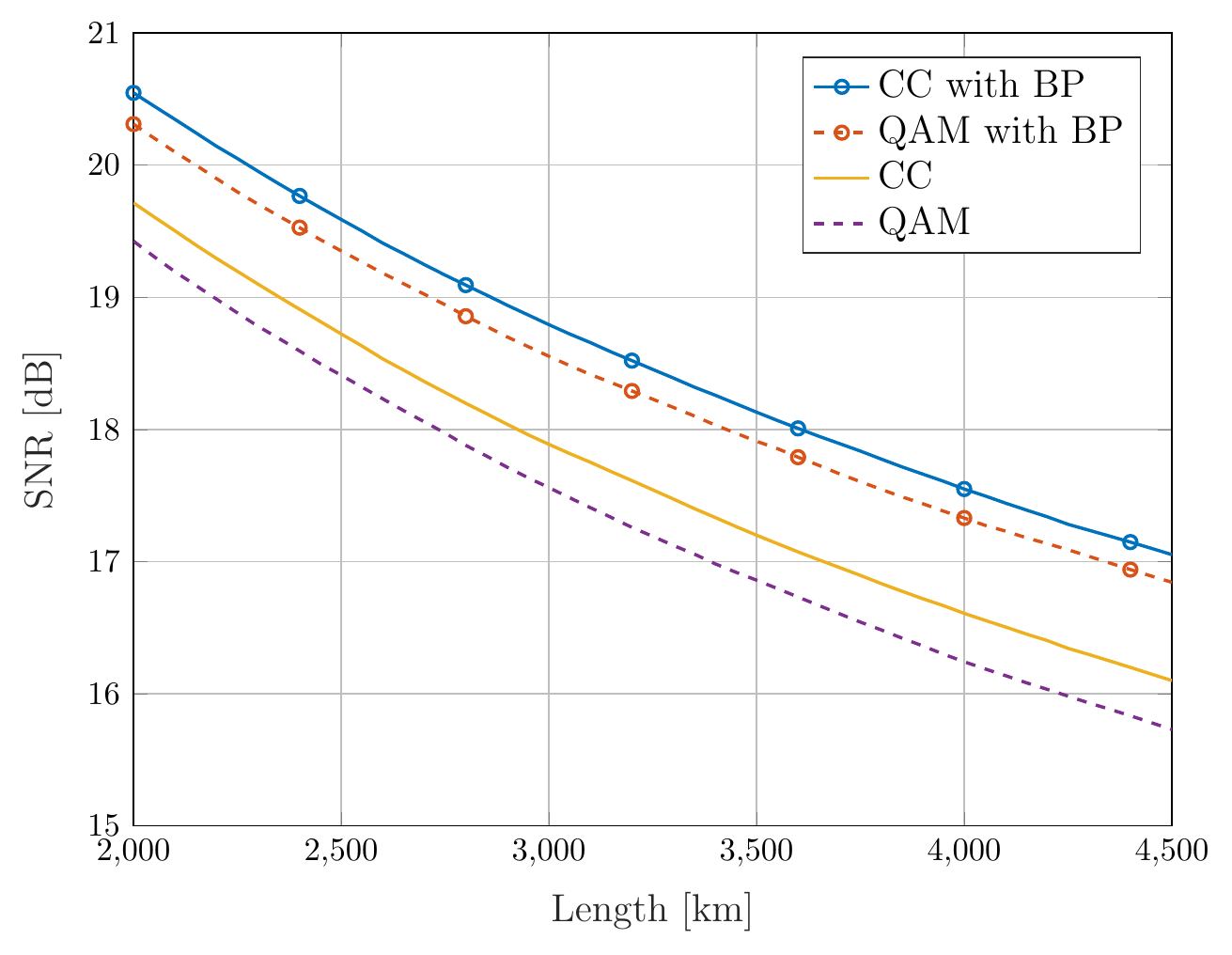}
\caption{SNR for a CC code of blocklength $171$ and IUD transmission using $64$-QAM versus fiber length. Two different detection schemes are considered: back-propagation (BP) followed by matched-filtering and matched-filtering without back-propagation.}
\label{fig:snr_noisy}
\end{figure}
%%%%%%%%%%%%%%%%%%%%%%%%%%%%%%%%%%%%%%%%%%%

The SNRs are shown in Fig.~\ref{fig:snr_nonoise_bps}. As expected, the SNR gain
of the CC code is considerably smaller when using BPS. The SNR gain without
back-propagation is reduced to $0.2$ to $0.5$~dB, while the SNR gain with
back-propagation is between $0$ to $0.5$~dB, depending on the length.

\subsection{More Realistic Experiments}

\begin{figure}[t]
\centering
\includegraphics[width=\columnwidth]{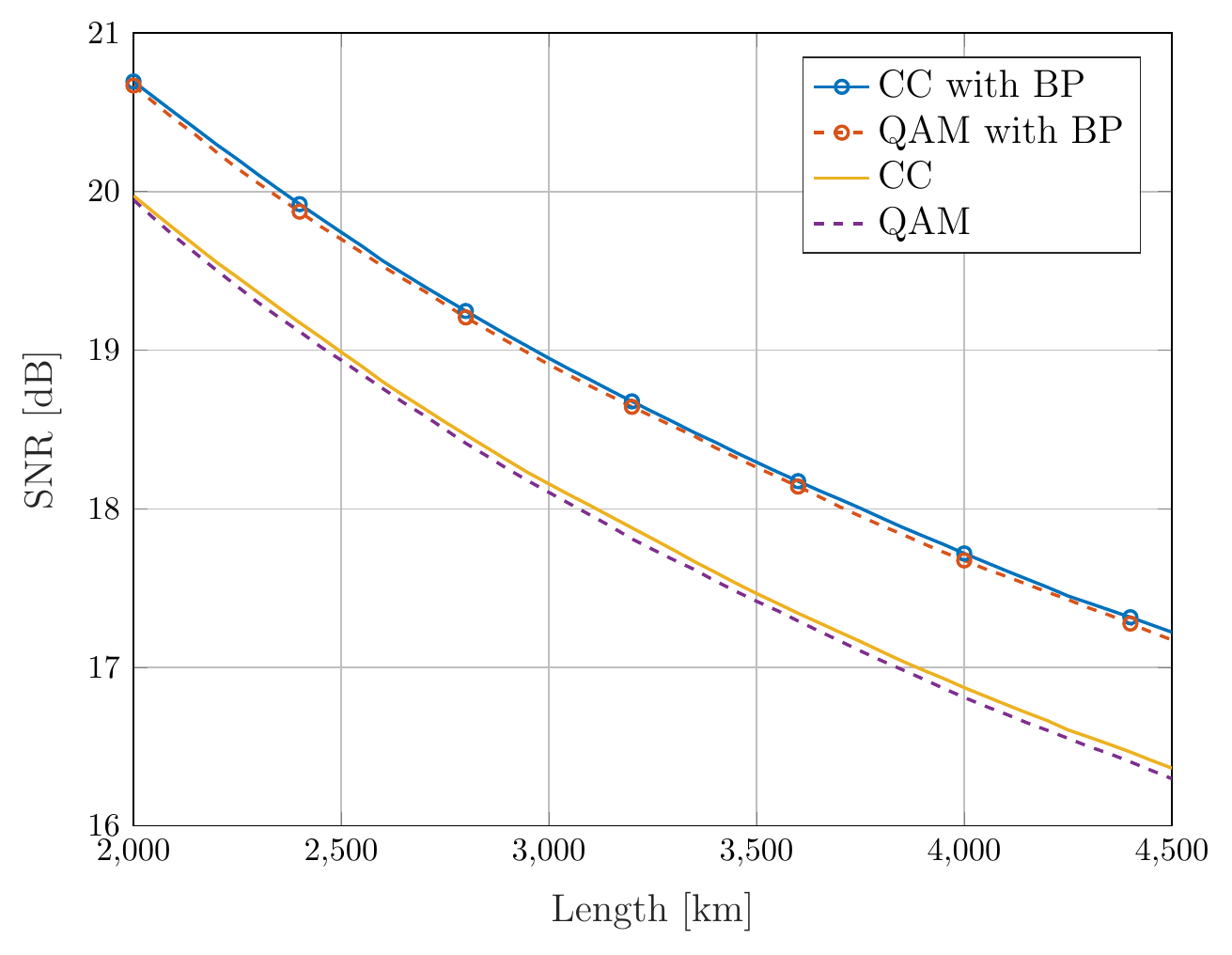}
\caption{SNR for a CC code of blocklength $171$ and IUD transmission using $64$-QAM versus fiber. Two different detection schemes are considered: back-propagation (BP) followed by matched-filtering and matched-filtering without back-propagation. The samples of the matched filter are back-rotated by using a genie-aided blind phase search algorithm.}
\label{fig:snr_noisy_bps}
\end{figure}

To properly simulate single-polarized data transmission over the optical fiber,
we take into account the attenuation of the fiber. It is assumed that an
erbium-doped fiber amplifier (EDFA) is located at the end of each span of
length $50$~km. Pulse shaping is done by using a root-raised cosine pulse with
a roll-off factor of about $6$\%. The channel spacing is $50$~GHz including
about $6$\% guard band. Among the five WDM channels considered, the channel of
interest, as before, is the middle one. The SNRs obtained for the CC code of
blocklength $171$ as well as the IUD transmission from a $64$-QAM constellation
are shown in Fig.~\ref{fig:snr_noisy} when no carrier recovery is in place. The
results of detection with BPS are shown in Fig.~\ref{fig:snr_noisy_bps}. The
SNR gain of CC codes in the absence of BPS is about $0.3$~dB without
back-propagation and about $0.2$~dB with back-propagation. With BPS, however,
the SNR gain of CC codes is negligible (about $0.05$~dB).

%%%%%%%%%%%%%%%%%%%%%%%%%%%%%%%%%%%%%%%%%%%
%%%%%%%%%%%%%%%%%%%%%%%%%%%%%%%%%%%%%%%%%%%

\section{Conclusions}\label{sec:conclusions}

Using a first order perturbation model derived
from the nonlinear Schr\"{o}dinger
equation, we have shown why short constant-composition codes reduce
nonlinear interference noise in a WDM system.  We have also shown that phase
tracking can be used to achieve the same reduction in nonlinear interference
noise even without using constant-composition codes. 

Our analysis shows that the more general class of spherical codes, which
encompass constant-composition codes, have the potential to reduce cross-phase
modulation, as long as their blocklength is not too long.  We leave the
investigation of the potential gains provided by spherical codes as future
work.

\IEEEtriggeratref{9}
\bibliographystyle{IEEEtran}
\bibliography{IEEEfull,mybib}

\begin{thebibliography}{10}
\providecommand{\url}[1]{#1}
\csname url@samestyle\endcsname
\providecommand{\newblock}{\relax}
\providecommand{\bibinfo}[2]{#2}
\providecommand{\BIBentrySTDinterwordspacing}{\spaceskip=0pt\relax}
\providecommand{\BIBentryALTinterwordstretchfactor}{4}
\providecommand{\BIBentryALTinterwordspacing}{\spaceskip=\fontdimen2\font plus
\BIBentryALTinterwordstretchfactor\fontdimen3\font minus
  \fontdimen4\font\relax}
\providecommand{\BIBforeignlanguage}[2]{{%
\expandafter\ifx\csname l@#1\endcsname\relax
\typeout{** WARNING: IEEEtran.bst: No hyphenation pattern has been}%
\typeout{** loaded for the language `#1'. Using the pattern for}%
\typeout{** the default language instead.}%
\else
\language=\csname l@#1\endcsname
\fi
#2}}
\providecommand{\BIBdecl}{\relax}
\BIBdecl

\bibitem{essiambre2010capacity}
R.-J. Essiambre, G.~Kramer, P.~J. Winzer, G.~J. Foschini, and B.~Goebel,
  ``Capacity limits of optical fiber networks,'' \emph{J.\ Lightw.\ Techn.},
  vol.~28, no.~4, pp. 662--701, Feb. 2010.

\bibitem{mecozzi2012nonlinear}
A.~Mecozzi and R.-J. Essiambre, ``Nonlinear {S}hannon limit in pseudolinear
  coherent systems,'' \emph{J.\ Lightw.\ Techn.}, vol.~30, no.~12, pp.
  2011--2024, Jun. 2012.

\bibitem{dar2017nonlinear}
R.~Dar and P.~J. Winzer, ``Nonlinear interference mitigation: Methods and
  potential gain,'' \emph{J.\ Lightw.\ Techn.}, vol.~35, no.~4, pp. 903--930,
  Feb. 2017.

\bibitem{amari2019introducing}
A.~Amari, S.~Goossens, Y.~C. G\"{u}ltekin, O.~Vassilieva, I.~Kim, T.~Ikeuchi,
  C.~M. Okonkwo, F.~M.~J. Willems, and A.~Alvarado, ``Introducing enumerative
  sphere shaping for optical communication systems with short blocklengths,''
  \emph{J.\ Lightw.\ Techn.}, vol.~37, no.~23, pp. 5926--5936, Dec. 2019.

\bibitem{fehenberger2020mitigating}
T.~Fehenberger, H.~Griesser, and J.-P. Elbers, ``Mitigating fiber
  nonlinearities by short-length probabilistic shaping,'' in \emph{Optical
  Fiber Commun. Conf.}\hskip 1em plus 0.5em minus 0.4em\relax {OSA}, 2020, p.
  Th1I.2.

\bibitem{fehenberger2020analysis}
T.~Fehenberger, D.~S. Millar, T.~Koike-Akino, K.~Kojima, K.~Parsons, and
  H.~Griesser, ``Analysis of nonlinear fiber interactions for finite-length
  constant-composition sequences,'' \emph{J.\ Lightw.\ Techn.}, vol.~38, no.~2,
  pp. 457--465, Jan. 2020.

\bibitem{peng2021baud}
W.-R. Peng, A.~Li, Q.~Guo, Y.~Cui, and Y.~Bai, ``Baud rate and shaping
  blocklength effects on the nonlinear performance of super-symbol
  transmission,'' \emph{Optics Exp.}, vol.~29, no.~2, pp. 1977--1990, Jan.
  2021.

\bibitem{wu2021temporal}
K.~Wu, G.~Liga, A.~Sheikh, F.~M.~J. Willems, and A.~Alvarado, ``Temporal energy
  analysis of symbol sequences for fiber nonlinear interference modelling via
  energy dispersion index,'' \emph{J.\ Lightw.\ Techn.}, vol.~39, no.~18, pp.
  5766--5782, Sep. 2021.

\bibitem{wickham2004bit}
L.~K. Wickham, R.-J. Essiambre, A.~H. Gnauck, P.~J. Winzer, and A.~R.
  Chraplyvy, ``Bit pattern length dependence of intrachannel nonlinearities in
  pseudolinear transmission,'' \emph{IEEE Photonics Techn.\ Lett.}, vol.~16,
  no.~6, pp. 1591--1593, Jun. 2004.

\bibitem{civelli2020interplay}
S.~Civelli, E.~Forestieri, and M.~Secondini, ``Interplay of probabilistic
  shaping and carrier phase recovery for nonlinearity mitigation,'' in
  \emph{Europ.\ Conf.\ Opt.\ Comm.}\hskip 1em plus 0.5em minus 0.4em\relax
  {IEEE}, 2020, pp. 1--4.

\bibitem{govind2017nonlinear}
G.~Agrawal, \emph{Nonlinear Fiber Optics}, 5th~ed.\hskip 1em plus 0.5em minus
  0.4em\relax Elsevier Academic Press, 2017.

\bibitem{rowe2015galsim}
B.~T. Rowe, M.~Jarvis, R.~Mandelbaum, G.~M. Bernstein, J.~Bosch, M.~Simet,
  J.~E. Meyers, T.~Kacprzak, R.~Nakajima, J.~Zuntz \emph{et~al.}, ``{GALSIM}:
  The modular galaxy image simulation toolkit,'' \emph{Astronomy and
  Computing}, vol.~10, pp. 121--150, Apr. 2015.

\bibitem{macleod1996rational}
A.~J. MacLeod, ``Rational approximations, software and test methods for sine
  and cosine integrals,'' \emph{Numerical Algorithms}, vol.~12, no.~2, pp.
  259--272, Sep. 1996.

\bibitem{sinkin2003optimization}
O.~V. Sinkin, R.~Holzlohner, J.~Zweck, and C.~R. Menyuk, ``Optimization of the
  split-step {F}ourier method in modeling optical-fiber communications
  systems,'' \emph{J.\ Lightw.\ Techn.}, vol.~21, no.~1, pp. 61--68, 2003.

\bibitem{pfau2009hardware}
T.~Pfau, S.~Hoffmann, and R.~No{\'e}, ``Hardware-efficient coherent digital
  receiver concept with feedforward carrier recovery for {$ M $-QAM}
  constellations,'' \emph{J.\ Lightw.\ Techn.}, vol.~27, no.~8, pp. 989--999,
  2009.

\end{thebibliography}

\end{document}